\documentclass[aps,prb,twocolumn,10pt]{revtex4-2}
\usepackage{amsmath}
\usepackage{textcomp}
\usepackage{graphicx}
\usepackage{rotating}
\makeatletter
\let\saved@includegraphics\includegraphics
\AtBeginDocument{\let\includegraphics\saved@includegraphics}
\renewenvironment*{figure}{\@float{figure}}{\end@float}

\usepackage{xcolor}
\usepackage[normalem]{ulem}

\begin{document}

\title{Tunable coupling of two mechanical resonators by a graphene membrane}

\author
{G.~J.~Verbiest$^{1,2,+,*}$, M.~Goldsche$^{1,2}$, J.~Sonntag$^{1,2}$, T.~Khodkov$^{1,2}$, N.~von~den~Driesch$^2$, \\
D.~Buca$^2$, and C. Stampfer$^{1,2}$
\\
%
\normalsize{$^1$JARA-FIT and 2nd Institute of Physics, RWTH Aachen University, 52074 Aachen, Germany, EU}\\
\normalsize{$^2$Peter Gr\"unberg Institute (PGI-8/9), Forschungszentrum J\"ulich, 52425 J\"ulich, Germany, EU}\\
\normalsize{$^+$Current address: Department of Precision and Microsystems Engineering, Delft University of Technology, Mekelweg 2, 2628 CD Delft, The Netherlands, EU}
\\
\normalsize{$^{*}$Corresponding author; E-mail: G.J.Verbiest@tudelft.nl}
}

\begin{abstract}
Coupled nanomechanical resonators are interesting for both fundamental studies and practical applications as they offer rich and tunable oscillation dynamics. 
At present, the mechanical coupling in such systems is often mediated by a fixed geometry, such as a joint clamping point of the resonators or a displacement-dependent force.
Here we show a graphene-integrated electromechanical system consisting of two physically separated mechanical resonators -- a comb-drive actuator and a suspended silicon beam -- that are tunably coupled by a graphene membrane. 
The graphene membrane, moreover, provides a sensitive electrical read-out for the two resonating systems silicon structures showing 16 different modes in the frequency range from 0.4~to 24~MHz. In addition, by pulling on the graphene membrane with an electrostatic potential applied to one of the  silicon resonators, we control the mechanical coupling, quantified by the $g$-factor, from 20 kHz to 100 kHz. Our results pave the way for coupled nanoelectromechanical systems requiring controllable mechanically coupled resonators.
\end{abstract}

\maketitle

Resonating silicon-based micro- and nanoelectromechanical systems can operate over a wide range of frequencies, varying from the kHz to the GHz regime, very much depending on the applications.
This includes high-quality-factor band pass filters \cite{yang2001,gouttenoire2010,piekarski2001}, signal amplifiers \cite{mathew2016dynamical,karabalin2011}, high-precision sensors (incl. biosensors) \cite{bogue2013}, or even logic gates \cite{tsai2008}. 
Moreover, mechanically coupled resonators have attracted increasing attention thanks to their interesting dynamics \cite{hajime2013substrate,luo2018graphene,deng2016nanotube,verbiest2016tunable}, improved performance and advanced tunability compared to single resonators~\cite{teufel2011coupling}. 
The mechanical coupling between different resonators can be well-designed~\cite{singh2018} and can be used e.g. as low-noise signal amplifier~\cite{singh2019}.
Yet, up to now, the coupling is mediated by a fixed geometric contact or clamping between the mechanical resonators or a position-dependent force, which limits the control over the coupling~\cite{okamoto2013}.
The implementation of an integrated and independent control of the mechanical coupling is still a major technological challenge.
This is mainly a consequence of the missing frequency tunability of the constituent mechanical resonators and their weak vibration coupling~\cite{okamoto2013}.

\begin{figure*}[t]
\begin{center}
\includegraphics[draft=false,keepaspectratio=true,clip,width=0.98\linewidth]{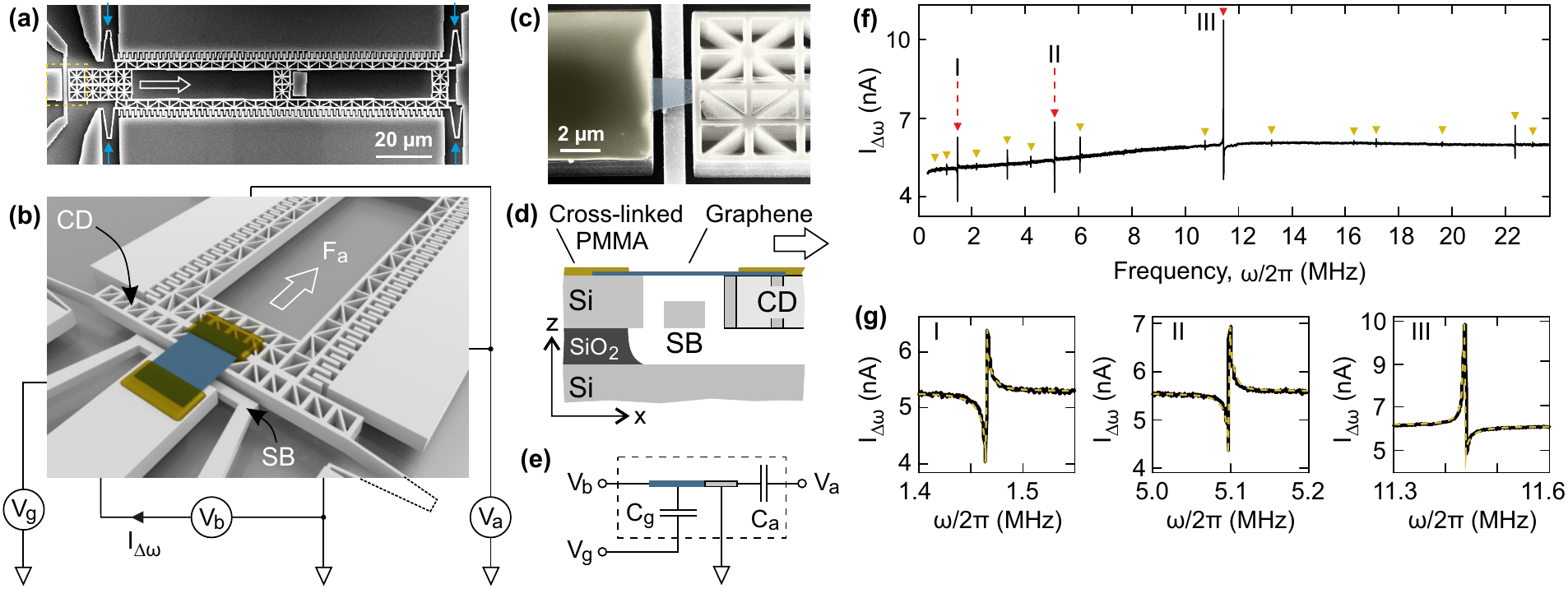} 
\end{center}
\caption{(a) Top view scanning electron microscope (SEM) image of a comb-drive (CD) actuator device. The suspended part of the CD actuator is held by four springs (blue arrows). (b) Illustration of an actuator device with integrated graphene (blue) clamped by cross-linked PMMA (yellow) and electrical contacts. The potential $V_\text{a}$ generates an electrostatic force $F_\text{a}$ in the direction of the white arrow. The potential $V_\text{g}$ tunes the charge carrier density in the suspended graphene membrane and $V_\text{b}$ is the applied bias, which results in a down-mixing current $I_{\Delta\omega}$ (see text). (c) False-color SEM image (taken under an angle) of the area (see dashed box in panel a) containing the graphene membrane (light blue). (d) Schematic cross-sectional view of the device, highlighting the graphene membrane the suspended silicon beam (SB) and the CD actuator. (e) A simplified electrical circuit diagram of the nanoelectromechanical device. (f) $I_{\Delta\omega}$ as a function of $\omega/2\pi$ showing sixteen mechanical resonances. This data was recorded at a temperature of $\sim2.3$~K and at a low $^4$He pressure (1~mbar). (g) Close-ups of the data in panel (f) around the main resonances: I (1.48 MHz), II (5.14 MHz) and III (11.44 MHz). The yellow dashed lines are fits to the data.
\label{fig1}}
\end{figure*}

Graphene-integrated nano-electromechanical systems are promising candidates for overcoming these limitations of silicon-based micro-electromechanical systems (MEMS). Graphene, an atomically thin crystal of carbon atoms, features a high mechanical strength \cite{lee2008measurement,tomori2011}, an unprecedented high carrier mobility \cite{mayorov2011}, and an highly sensitive electrical read-out scheme for its mechanical motion \cite{chen2013graphene,katsnelson2007,juan2013}.
Moreover, the low mass density and the high Young's modulus \cite{Novoselov2004,lee2008measurement} make graphene interesting for resonator based sensor applications \cite{lemme2020}, for example as force \cite{mashoff2010,chen2013review,chaste2012nanomechanical}, ultrasound \cite{verbiest2018graphene,laitinen2019,todorovic2015,xu2010radio} or pressure sensors \cite{dolleman2016}. 
There is also a growing interest to implement graphene as a mechanical element in silicon MEMS devices such as accelerometers \cite{hurst2015,fan2019}, since their high flexiblity allows for a considerable scaling down of the device footprint while maintaining high sensitivity. These prototype demonstrations show that graphene is an interesting candidate for the integration in MEMS as a motion sensor of spring. Despite its high flexibility and the large tuning range of stiffness, the implementation of graphene as a tunable spring and mechanical coupler in silicon based MEMS devices has up to now not been demonstrated.

Here we show that a suspended graphene membrane can be used to couple two physically separated mechanical resonators. Moreover, the graphene membrane simultaneously provides an electrical read-out scheme for the motion of both of these resonators. The mechanical coupling between the resonators, mediated by the graphene membrane, can be controlled over a wide range by an electrostatic potential, realizing substantially enhanced coupling, when compared to systems without integrated membranes \cite{okamoto2013}.


The device was fabricated by an electron-beam lithography (EBL) based structuring of a Cr/Au/Cr hard mask on a silicon-on-insulator substrate consisting of 725 $\mu$m silicon, 1 $\mu$m SiO$_\text{2}$ and 2 $\mu$m highly p-doped silicon followed by a deep reactive ion etching (DRIE) step, as described in detail in Refs. \cite{goldsche2018tailoring,goldsche2018fabrication}.
The silicon beam (SB), which also functions as a bottom electrostatic gate for tuning the graphene properties, was fabricated by interrupting the DRIE step after etching 275~nm deep followed by the deposition of an additional Cr mask before etching completely through the highly p-doped silicon layer.
After removal of the Cr, a graphene/PMMA stack is transferred on the patterned comb-drive (CD) actuator.
Raman spectroscopy confirms the single-layer nature of the graphene flake (Supplementary Figure~1).
By an additional EBL step we partly cross-link the PMMA to clamp the graphene membrane onto the actuator on one side and to a fixed anchor on the other side (Figures \ref{fig1}a to \ref{fig1}d).
Finally, the actuator with the integrated graphene membrane is released from the substrate by removing the SiO$_\text{2}$ layer with 10\% hydrofluoric (HF) acid solution followed by a critical point drying (CPD) step. In the measurements presented here, the suspended graphene membrane has a length of $L \approx 2$~$\mu$m and a width of $W \approx 3$~$\mu$m (Supplementary Figure~1). 
The measurements were performed in a $^3$He/$^4$He dilution refrigerator with a base temperature around 20~mK, unless otherwise stated.

Figures~\ref{fig1}b and \ref{fig1}e depicts the electrical scheme of the measured device. A potential difference $V_\text{a}$ between the asymmetrically placed fingers of the CD actuator gives rise to an electrostatic force, $F_\text{a} = \tfrac{1}{2} \partial_x C_\text{a} V_\text{a}^2$ that pulls the suspended comb in the $x$-direction (see Fig.~\ref{fig1}d and white arrow in Fig.~1b). Here, the capacitance $C_\text{a}$ and $\partial_x C_\text{a}$ are given by the zeroth and the first order term in the displacement $\delta x$ of the actuator in a series expansion of the parallel plate approximation for the capacitance between its fingers \cite{goldsche2018tailoring}. 

\begin{figure*}[t]
\begin{center}
\includegraphics[draft=false,keepaspectratio=true,clip,width=1\linewidth]{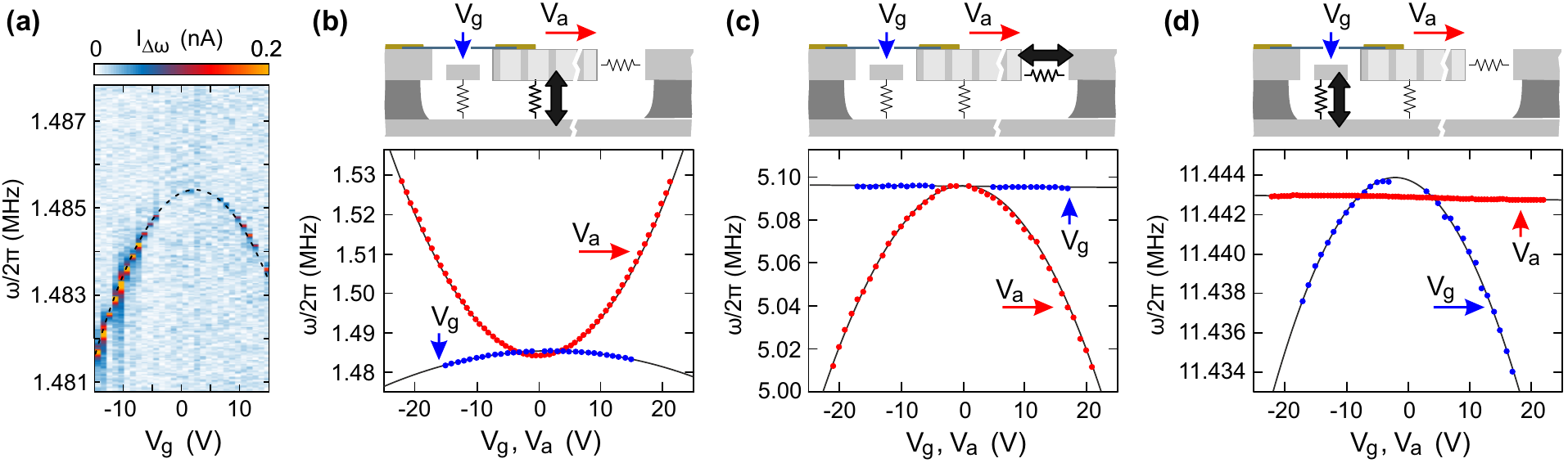}
\end{center}
\caption{(a) Down-mixing current $I_{\Delta\omega}$ as a function of $\omega/2\pi$ and $V_\text{g}$ (raw data) The dashed line highlights the resonance frequency. (b-d bottom) Frequency dependence of the resonances I, II, and III on the applied potentials $V_\text{g}$ (blue) and $V_\text{a}$ (red). The black lines are quadratic fits in accordance to the applied electrostatic force (see text). (b-d top) Schematic illustrations of the mechanical system. The black arrows indicate the main vibrating component of the corresponding resonance. The red and blue arrow indicate the direction of the force induced by $V_\text{g}^2$ and $V_\text{a}^2$, respectively. 
\label{fig2}}
\end{figure*}

\begin{figure*}[t]
\begin{center}
\includegraphics[draft=false,keepaspectratio=true,clip,width=1\linewidth]{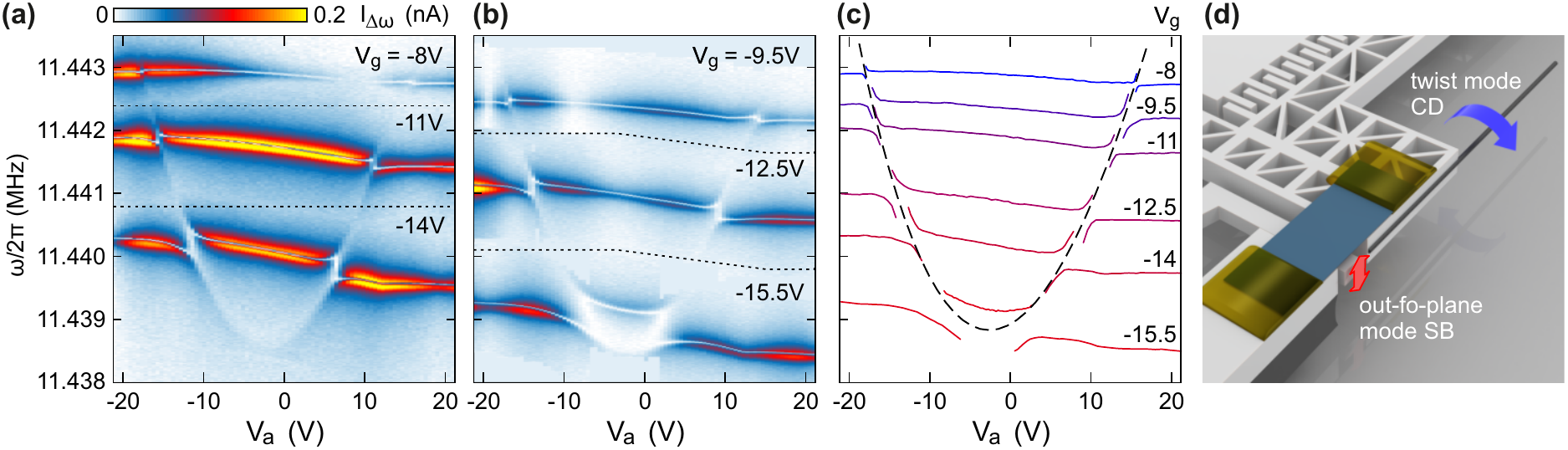}
\end{center}
\caption{(a-b) Raw data of the down-mixing current $I_{\Delta\omega}$ as a function of $\omega/2\pi$ and $V_\text{a}$ for different fixed $V_\text{g}$ (black labels) reveal avoided crossings. Different measurements are stitched together at the dotted lines. (c) The colored lines indicate the extracted resonance frequencies for different fixed $V_\text{g}$ (black labels). The dashed black line traces the dependence of the avoided crossings on $V_\text{a}$. (d) Schematic illustration highlighting the two interacting modes.
\label{fig3}}
\end{figure*}

An AC potential $V_\text{g}^\text{ac}$ on the suspended SB at frequency $\omega/2\pi$ supplies an external force acting on the suspended graphene membrane. 
The mechanical displacement of the graphene $\delta z$ perpendicular to the membrane plane ($z$-direction, see Figure \ref{fig1}d) at frequency $\omega/2\pi$ modulates its conductance $G$. The other (in-plane) directions do not modulate the conductance $G$ and thus do not contribute to the signal \cite{chen2013graphene}. As $\omega/2\pi$ is usually in the MHz range, a drain-source bias $V_\text{b}$ is applied at a slightly different frequency $(\omega\pm\Delta\omega)/2\pi$ to generate a down-mixed current $I_{\Delta\omega}$ passing the graphene membrane at a low, measurable frequency $\Delta\omega/2\pi$ \cite{chen2009performance}:
\begin{equation}
I_{\Delta\omega} = \frac{1}{2} V_\text{b} \frac{\partial G}{\partial V_\text{g}} \left( V_\text{g}^\text{ac} + V_\text{g} \frac{\partial_z C_\text{g}}{C_\text{g}} \delta z \right),
\label{eq1}
\end{equation}
\noindent
where $V_\text{g}$ is the applied DC potential on the SB acting as gate, and $\partial G / \partial V_g$ is the transconductance (Supplementary Figure~2).
The capacitance $C_\text{g} = 0.19$~fF and $\partial_z C_\text{g} = -0.70$~nF/m are given by the zeroth and the first order term in $\delta z$ of an analytical series expansion of the parallel plate approximation for the capacitance between the suspended SB and the graphene membrane.

To extract the resonance frequencies of the device, we measure $I_{\Delta\omega}$ as a function of $\omega/2\pi$.
The mechanical resonances are observed as dips and peaks in $I_{\Delta\omega}$ (see Figure~\ref{fig1}f).
We observe in total sixteen resonances in the range from $0.4$ to $24$~MHz.
In this work, we focus on the three main resonances (labelled as I, II, and III).
Corresponding close-ups are shown in Figure \ref{fig1}g. We fit the resonances with a nonzero-phase Lorentzian \cite{chen2009performance} to extract the resonance frequency $\omega_0/2\pi$, the quality factor $Q$, and the effective drive amplitude $A$.

To understand the physical origin of the different resonances, we extract the effective masses and spring constants of the resonances from the tuning of the resonance frequencies with applied electrostatic potentials.
Figure~\ref{fig2}a shows the measured down-mixed current as a function of $V_\text{g}$ for $V_\text{a} = 0$ V. From such data, we extract the dependencies of the resonance frequencies on the applied potentials.
Figures~\ref{fig2}b-d show the dependencies of resonances I, II, and III on $V_\text{g}$ (blue) and $V_\text{a}$ (red).
Resonances I and III tune towards lower frequencies for increasing $|V_\text{g}|$, whereas resonance II tunes towards lower frequencies for increasing $|V_\text{a}|$.
The tuning towards lower frequencies for increasing $|V_\text{g}|$ or $|V_\text{a}|$ suggests a dominating capacitive softening effect~\cite{wu2011capacitive,eichler2011nonlinear,song2011stamp,kozinsky2006tuning}.
Therefore, we fit the tuning of $\omega_0$ with \cite{chen2009performance}
\begin{equation}
\omega_0 = \sqrt{\frac{k_0-\frac{1}{2} \partial^2_{x(z)} C_\text{a(g)} V_\text{a(g)}^2}{m_\text{eff}}},
\label{eq2}
\end{equation}
\noindent
where $k_0$ is the effective spring constant, $m_\text{eff}$ is the effective mass, $\partial^2_x C_\text{a} = 97$ mF/m$^\text{2}$ characterizes the capacitive softening of the actuator, and $\partial^2_z C_\text{g} = 5.1$ mF/m$^\text{2}$ characterizes the capacitive softening of the graphene-SB capacitance $C_\text{g}$ (see Figure 1e). 
Here, $\partial^2_{x(z)} C_\text{a(g)}$ denotes the second order term in $\delta x(z)$ of an analytical series expansion of the parallel plate approximation for $C_\text{a(g)}$.
The fit results for the resonances I, II, and III are depicted by black lines in the lower panels of Figures~\ref{fig2}b-d. The fit parameters are summarized in Supplementary Table~1.
Resonances I and II have an effective mass $m_{\text{eff}}$ of $1.75 \pm 0.02$~ng and $0.56 \pm 0.01$ ng, respectively, and that of resonance III is $0.065 \pm 0.001$~ng. We confirmed the extracted values for $m_\text{eff}$ (and $k_0$) with a second independent measurement, in which we extracted the effective drive amplitude $A$ as a function of the driving force \cite{chen2009performance,sazonova2004tunable} $F_{d} = \partial_z C_\text{g} V_\text{g} V_\text{g}^\text{ac}$ by varying $V_\text{g}^{ac}$ for a fixed $V_\text{g}$ and $V_\text{a}$ (Supplementary Figure~3).
In this measurement, the measured transconductance $\partial G / \partial V_g$ (Supplementary Figure~2), in combination with the applied potentials and the estimated capacitances allows us to quantitatively extract the physical vibration amplitudes $A$ contained in $\delta z$.
The effective masses for our device are at least three orders of magnitude larger than the ones typically observed for graphene resonators~\cite{chen2009performance}, indicating that the graphene mass is irrelevant for the total device.

The effective masses of resonances I and II are comparable to the estimated mass of the CD actuator when taking a density of 2329 kg/m$^\text{3}$ for the highly p-doped silicon leading to $m_\text{CD} = 1.79$~ng, in good agreement with the effective mass extracted from resonance I. The effective mass extracted from resonance II is roughly one-third of $m_\text{CD}$, which can be explained by the effective mass reduction for a doubly clamped beam \cite{hauer2013} in its fundamental mode and thus highlights the importance of the mode shape. The effective mass $m_{\text{eff}}$ extracted from resonance III is approximately equal to the estimated mass of the suspended silicon beam, $m_\text{SB} = 0.048$~ng.  This suggests that the observed resonances can be attributed to resonances of the actuator and of the silicon beam, and not to the mechanical motion of the graphene membrane.

To verify the origin of the resonances and to clarify the mode shapes, we performed finite element calculations \cite{Comsol}. It is important to include the suspended graphene membrane in the simulations for two reasons: (i) we can only measure modes with an oscillation of the graphene membrane in the $z$-direction (see equation~\ref{eq1}) and (ii) the spring constant $\sim 6.7$~N/m of the silicon actuator in the $x$-direction is much smaller than the expected spring constant $k_\text{gr} \sim Y_\text{2D} W/L = 510$~N/m of the graphene membrane. 
Here we used the literature value $Y_\text{2D} = 340$~N/m for the two-dimensional Young's modulus of graphene that is expected at cryogenic temperatures \cite{lee2008measurement,liu2016mechanical,nicholl2015}.
The highly p-doped silicon has a Young's modulus of $\approx $160 GPa \cite{hopcroft2010young,li2003ultrathin}.
We find excellent agreement between all the measured and computed resonance frequencies: the ratio between them is on average $1.04 \pm 0.06$ (see Supplementary Table~2).
The top panels in Figures~\ref{fig2}b-d schematically illustrate the main vibrating components for resonances I, II, and III. We find that resonances I and II correspond to an out-of-plane and in-plane motion of the CD actuator. Resonance III is an out-of-plane mode of the silicon beam, which is supported by the absence of any tunablility with $V_\text{a}$.
Details of all computed frequencies and mode shapes are provided in Supplementary Table~2 (and Supplementary Figure~4).
The computed mode shapes are consistent with the observed capacitive softening.
In total we directly detect sixteen mechanical resonances of the CD actuator and one of the suspended silicon beam.
As the observed frequency-tuning is in agreement with capacitive softening, we can use the extracted effective spring constants to determine the static displacements $\delta x$ and $\delta z$. We compute these displacements by dividing the electrostatic forces $F_\text{a} = \tfrac{1}{2} \partial_x C_\text{a} V_\text{a}^2$ and $F_\text{g} = \tfrac{1}{2} \partial_z C_\text{g} V_\text{g}^2$ by the spring constant of the lowest in-plane and out-of-plane mode, respectively (Supplementary Figure~5).
The in-plane displacement $\delta x$ goes up to 3~nm and is in agreement with the related strain values extracted by Raman spectroscopy measurements on similar devices~\cite{goldsche2018tailoring}.

\begin{figure*}[t]
\begin{center}
\includegraphics[draft=false,keepaspectratio=true,clip,width=1\linewidth]{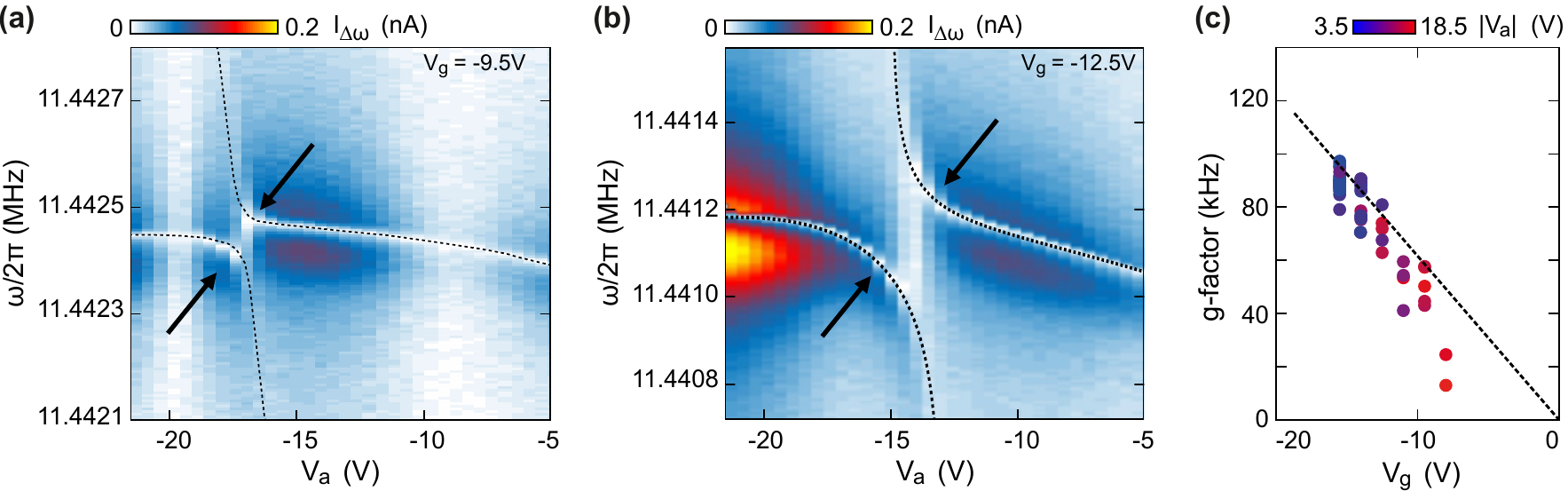} 
\end{center}
\caption{(a)-(b) Zoom of avoided crossings. The $g$-factor is proportional to the minimal distance between the resonance lines (dashed black lines). (c) The $g$-factor extracted from the avoided crossings as a function of $V_\text{g}$ reveals a tunable mode coupling. The dashed black line is the $g$-factor arising from the electrostatic force.
\label{fig4}}
\end{figure*}

\begin{figure}[h]
\begin{center}
\includegraphics[draft=false,keepaspectratio=true,clip,width=85mm]{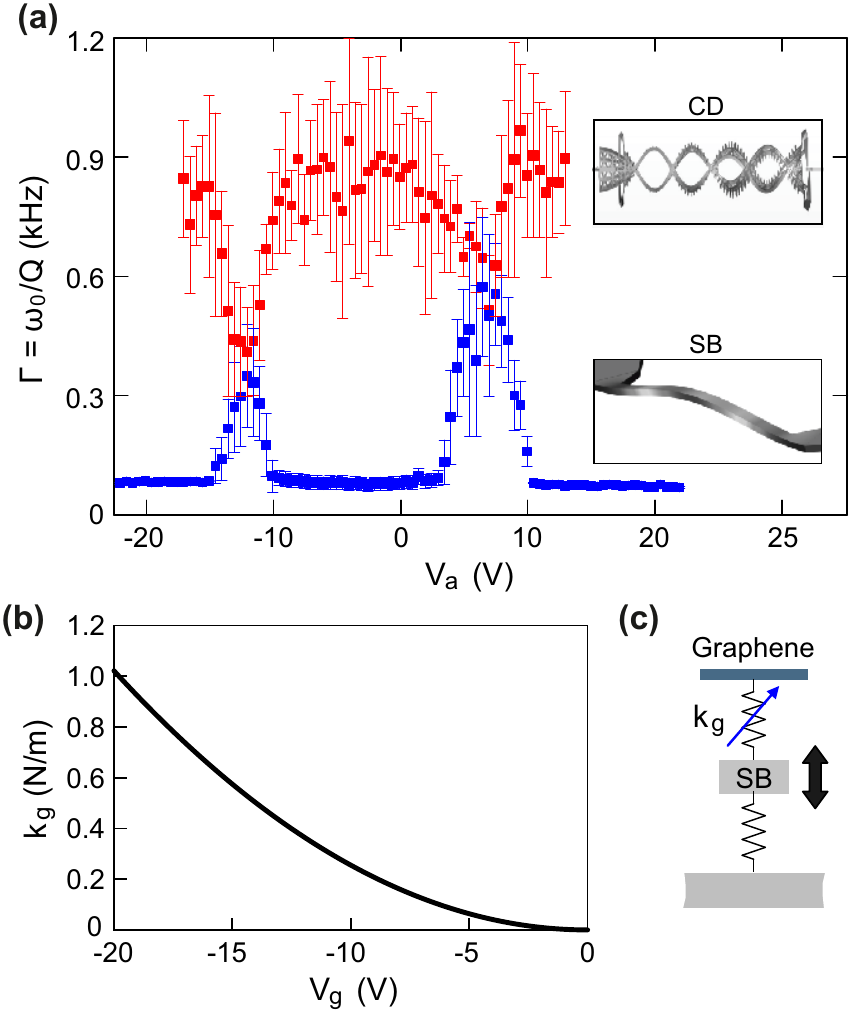} 
\end{center}
\caption{(a) Linewidths $\Gamma = \omega_0/Q$ of the interacting modes at $V_\text{g} = -14$ V. The linewidths of the two modes (see insets and Supplementary Figure~4) become equal near the avoided crossing. (b) The tunable electrostatic softening spring $k_\text{g}$ as a function of $V_\text{g}$. (c) Schematic illustration showing how $k_\text{g}$ connects the graphene membrane to the suspended silicon beam.
\label{fig5}}
\end{figure}

To understand the increase of resonance I with applied $|V_\text{a}|$ in Figure~\ref{fig2}b, we now focus on the graphene membrane.
As illustrated in the top panel of Fig.~\ref{fig2}b, resonance I is dominated by a spring along the $z$-direction, i.e. by an out-of-plane motion of the CD. Both the actuator and the graphene contribute to this spring.
The contribution of the graphene membrane 
is known to be highly sensitive to the induced strain $\Delta\epsilon$, and thus to a $\delta x$ displacement \cite{lau2012properties,lopez2017influence}.
As the total spring constant is known from the resonance itself, and the in-plane displacement of the actuator as well as the dimensions of the graphene membrane are known, the only free parameter is the Young's modulus of the graphene membrane (see Methods).
Requiring the same in-plane actuator displacement for tuning resonance I as for tuning resonance II, we obtain an effective Young's modulus of $Y_\text{2D} = 350\pm20$ N/m (see Methods and Supplementary Figure~5), which is in good agreement with values reported in the literature \cite{lee2008measurement,liu2016mechanical} and supports the used value in the finite element calculations.
This type of analysis gives us a complete understanding of the mechanical behavior of the system. 

Let us next focus on the resonance III attributed to the suspended silicon beam.
When measuring this resonance as a function of $V_\text{g}$ and $V_\text{a}$, we observe the emergence of avoided crossings, which is a clear signature for two strongly coupled modes. Figures~\ref{fig3}a and \ref{fig3}b show the down-mixing current as function of $V_\text{a}$ for various $V_\text{g}$ values (separated by dashed lines). We fit these measurements to extract the resonance frequency as function of $V_\text{a}$ for various $V_\text{g}$ values, which are plotted in Figure~\ref{fig3}c. Remarkably, we only observe both interacting modes at the avoided crossings. This suggests that the graphene membrane has no measurable motion in the $z$-direction for the mode with which the silicon beam is interacting. However, we can reconstruct the dependence of the mode with which the silicon beam is interacting by tracing the position of the avoided crossing as a function of $V_\text{a}$, as indicated by the black dashed parabola in Figure~\ref{fig3}c (for more data see Supplementary Figure~6).
The computed resonance frequency spectrum of the CD actuator shows a twist mode close to the one of the suspended SB with negligible net graphene motion (displacement) in $z$-direction. We thus attribute the avoided crossing to a strong coupling between the silicon beam and the twist mode of the CD actuator (see illustration in Fig.~\ref{fig3}d).

%

We extract the coupling strength between the modes, i.e. the so-called $g$-factor~\cite{verbiest2016tunable}, $g=\omega_\text{sep}/2\pi$ and linewidth $\Gamma =\omega_0/2\pi Q$ from each individual avoided crossing (see close-ups in Figure~\ref{fig4}a and \ref{fig4}b).
Figure~\ref{fig4}c shows that the strong coupling ($g > \Gamma$) between the modes is tunable with applied $|V_\text{g}|$, i.e. by the electrostatic force between the silicon beam and the graphene membrane~\cite{verbiest2016tunable}. Figure~\ref{fig5}a shows $\Gamma$ as function of $V_\text{a}$ highlighting that both modes fully hybridize with equal energy transfer between them \cite{teufel2011coupling}, which is another sign of strong coupling.
The linewidth comparison was performed on the data at $V_\text{g} = -14$~V, which show the twist mode nearly over the full $V_\text{a}$-range.
The observed $g$-factor is in agreement with the coupling expected from the electrostatic softening between the silicon beam and the graphene membrane, as shown in Figure~\ref{fig5}b.
Both the motion of (i) the silicon beam and (ii) the graphene membrane alter their separation, resulting not only in a shift of the resonance frequency, i.e. the well-known electrostatic softening~\cite{chen2009performance}, but also in a change of the coupling between the two resonators. Following equation~(\ref{eq2}), this coupling can be characterized by an effective spring constant $k_\text{g} \approx \tfrac{1}{2} (\partial^2 C_\text{g} / \partial z^2) V_\text{g}^2$ (Figures~\ref{fig5}b,5c).
Then the $g$-factor can be estimated without making any assumptions on the mode shape by $g = \sqrt{k_\text{g} / m_\text{eff}}/2\pi$. Here, the effective mass $m_\text{eff}$ is the one of the hybridized modes. As the mass of the suspended silicon beam is much smaller than that of the twist mode, we set $m_\text{eff}$ equal to the mass $m_{\text{CD}}$ of the actuator. When using the estimated mass of the actuator ($m_{\text{CD}} = 1.79$~ng) for $m_\text{eff}$ as well as the value for $(\partial^2 C_\text{g} / \partial z^2) = 5.1$~ mF/m$^\text{2}$ given above, we find a remarkably good agreement with the experimentally extracted $g$-factor and the computed one (see dashed line in Figure~\ref{fig4}c). Thus, the suspended graphene membrane allows for a strong coupling between two physically separated resonators, and even allows for the tuning of this coupling by an applied voltage. 

In summary, by taking advantage of the high sensitivity of graphene resonators, we implement and quantitatively validate a new coupling scheme for nano-electromechanical systems.
The coupling strength can be tuned from 20~kHz to 100~kHz and be completely switched off by an electrostatic potential.
We thus realise a maximal coupling of almost 1000 times larger than in systems without any integrated graphene membranes \cite{okamoto2013} and approximately equal to that obtained for spatially separated graphene resonators \cite{luo2018graphene}.
The resonators themselves are not affected by the light-weighted graphene membrane.
This coupling scheme is on-chip, poses no restrictions on the choice of material for the connected masses. It is possibly scalable by means of integrated graphene obtained via chemical vapour deposition, and can be possibly extended using other conducting two-dimensional materials instead.
The presented technique provides a platform to study the route to chaos in nonlinear dynamics by systematically measuring the orbit diagram~\cite{Strogatz2001}.
Additionally, the presented scheme enables one to switch on and off the mechanical coupling, giving rise to read-out schemes for quantum states at a well defined time instant with minimal back-action effects at other times.

{\bf Methods.}
\emph{Young's modulus extraction.}
The effective spring constant $k_\text{eff}$ is given by the one of the actuator in parallel to the out-of-plane stiffness of the graphene membrane.
The out-of-plane stiffness of the graphene membrane depends on the pre-strain $\epsilon_0$ \cite{chen2009performance}.
For a fixed $V_\text{g}$, we find a direct relation between the strain induced by the actuator $\Delta\epsilon$ and the change of the out-of-plane spring constant $\Delta k$ of the graphene membrane:
\begin{equation}
\Delta k = \frac{16 Y_\text{2D} W}{3L} \Delta\epsilon.\label{eq3}
\end{equation}
\noindent
Here, $W$ ($L$) is the width (length) of the suspended graphene membrane.
Note that we neglect the increase in strain by pulling upon the graphene with $V_\text{g}$ for two reasons:
Firstly, $V_\text{g}$ is constant, and secondly, the in-plane spring constant of the CD actuator ($\sim 6.7$~N/m) is much smaller than the expected in-plane spring constant of the graphene membrance ($\sim 540$~N/m), thereby minimising the strain induced with $V_\text{g}$.
We extract $\Delta k$ and $\Delta\epsilon$ experimentally and determine $W$ and $L$ from optical and scanning electron microscope images, which leaves $Y_\text{2D}$ as the only free parameter.
We extract $\delta x$ from the in-plane modes of the CD actuator.
The induced strain is then simply $\Delta\epsilon = \delta x/L$.
We then determine $\Delta k$ from the observed increase in resonance frequency of the out-of-plane mode:
\begin{equation}
\Delta k = k_\text{eff}\left( \frac{\omega_0(V_\text{a})^2}{\omega_0(V_\text{a} = 0\,\text{V})^2} - 1\right).\label{eq4}
\end{equation}
In Supplementary Figure~5, we plot $3L \Delta k / 16 W$ as a function of $\Delta\epsilon$, such that the slope is directly providing $Y_\text{2D}$.


{\bf Data Availability}
The data that support the findings of this study are available from the corresponding author upon reasonable request.

{\bf Associated Content}
The Supporting Information is available free of charge on the ACS Publications website at DOI:

{\bf Author Information}

\emph{Author Contribution} 
GV and MG executed the experiments.
GV designed the experiments and analysed the data.
MG, JS, and TK fabricated the devices. 
NvdD and DB provided support in device fabrication.
CS supervised the overall project.
All authors contributed to writing and reviewing the paper.

\emph{Corresponding Author} 
Correspondence and requests for materials should be addressed to G.V.~(email: G.J.Verbiest@tudelft.nl).

\emph{Notes}
The authors declare that there are no competing interests.

{\bf Acknowledgements}
The authors thank F. Haupt for help on the manuscript, R. Dolleman and M. Siskins for proof-reading, and S.~Staacks for help on the figures.
Support by the ERC (GA-Nr. 280140), the Helmholtz Nanoelectronic
Facility (HNF) \cite{hnf2017} at the Forschungszentrum J\"ulich, and the Deutsche Forschungsgemeinschaft (DFG) (STA 1146/12-1) are gratefully acknowledged.
G.V. acknowledges funding by the Excellence Initiative of the German federal and state governments.


\bibliographystyle{naturemag}
\bibliography{reflist}

\renewcommand{\figurename}{Supplementary Figure}
\setcounter{figure}{0}

\newpage
\begin{figure*}
\begin{center}
\includegraphics[width=174.6mm]{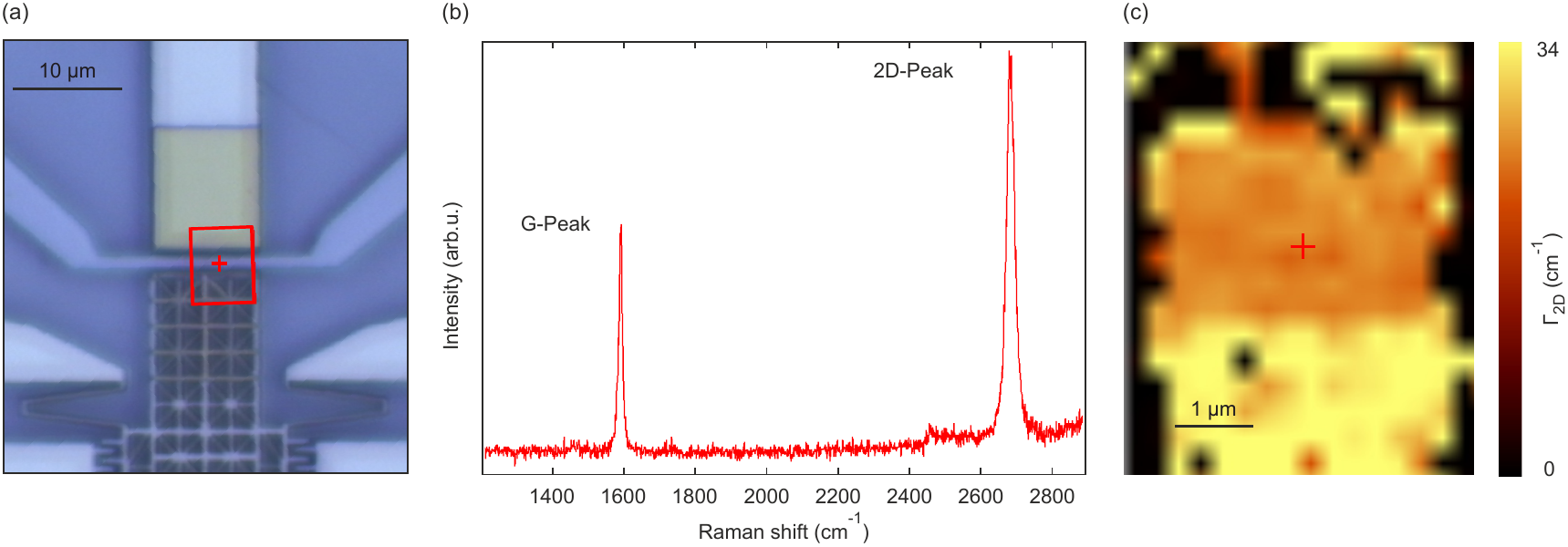}
\end{center}
\caption{
%
(a) Optical image of the region of the comb-drive actuator where the graphene membrane has been transfered. The graphene membrane is placed inside the red box.
(b) Raman spectrum of graphene measured at the position indicated with red cross in panel (a), showing the Raman G- and 2D-peak.
    The absence of a D-peak shows the good quality of the integrated graphene flake.
(c) Width of the Raman 2D-peak measured in the area outlined by the red box in panel (a).
    The average width of the Raman 2D-peak is 26.4 cm$^\text{-1}$, which shows single layer nature of the integrated graphene membrane \cite{hao2010}.
    The 2D-peak broadens or disappears in regions where cross-linked PMMA is present.
\label{S_fig_01}}
\end{figure*}

\begin{figure*}
\begin{center}
\includegraphics[width=121.9mm]{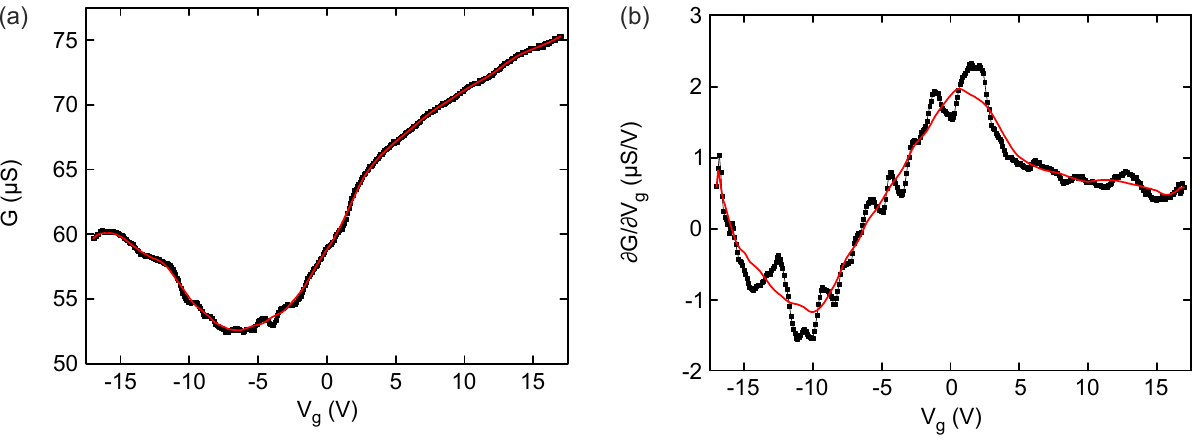}
\end{center}
\caption{
Measured conductance ($G$) and transconductance ($\partial G/\partial V_\text{g}$) of the integrated graphene membrane as a function of the potential $V_\text{g}$ applied on the suspended silicon beam.
(a) The conductance $G$ measured at 20~mK.
The extracted transconductance $\partial G/\partial V_\text{g}$ from the measurement in panel (a) is shown in panel (b).
The red lines show the smoothing splines of the (trans)conductance.
\label{S_fig_02}}
\end{figure*}

\begin{figure*}
\begin{center}
\includegraphics[width=125mm]{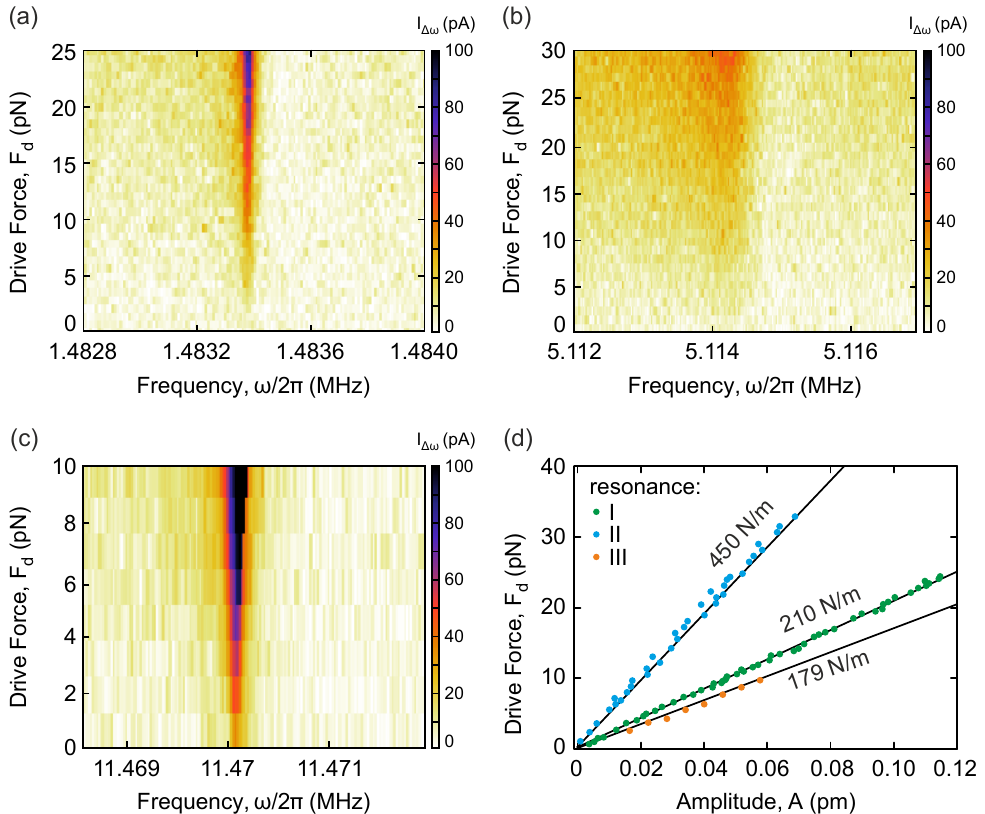}
\end{center}
\caption{
Drive amplitude as a function of drive force.
(a)-(c) Measured down-mixing current $I_{\Delta\omega}$ (see main manuscript) as a function of the drive force $F_{d} = \partial_z C_\text{g} V_\text{g} V_\text{g}^\text{ac}$ (for more details see main manuscript) and frequency $\omega/2\pi$ close to the resonances I, II, and III introduced in the main manuscript. By fitting each trace to a nonzero phase Lorentzian \cite{chen2009performance}, we find the effective drive amplitude $A$, which we convert into a physical amplitude in nanometer with equation~1 in the main text, using the known transconductance (see Supplementary~Figure~2).
(d) The drive force $F_\text{d}$ as function of the extracted drive amplitude $A$ for resonances I, II, and III. The slopes are equal to the effective spring constant $k_\text{eff}$ of the different resonances (see labels).
\label{S_fig_04}}
\end{figure*}

\begin{figure*}
\begin{center}
\includegraphics[width=\linewidth]{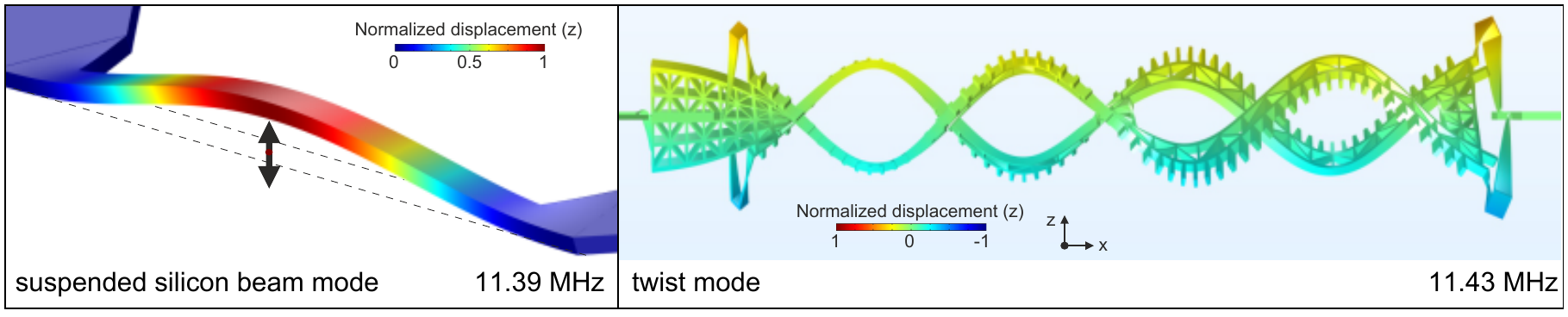}
\end{center}
\caption{
Mode shape of resonances III (left panel) and the twist mode of the comb-drive actuator with which it interacts (right panel). Both are extracted from a finite element analysis.
\label{S_fig_07}}
\end{figure*}

\begin{figure*}
\begin{center}
\includegraphics[width=125mm]{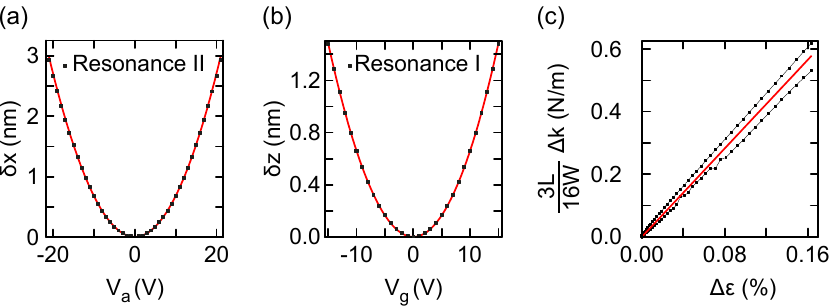}
\end{center}
\caption{
(a) Estimated $\delta x$ of the comb-drive actuator as a function of $V_\text{a}$ from resonance II. (b) Estimated $\delta z$ of the comb-drive actuator as a function of $V_\text{g}$ from resonances I. (c) The black data points represent the scaled change in spring constant $\tfrac{3L}{16W}\Delta k$ (see methods) of resonance I as a function of the strain $\Delta\epsilon$ induced with the comb-drive actuator. The strain $\Delta\epsilon$ is estimated from the data in panel (a) and $\tfrac{3L}{16W}\Delta k$ is computed from the change in resonance frequency. The black line connects the data points and serves as a guide to the eye. The red line represents a linear fit to the black data points. The slope of the red line equals the Young's modulus and has a value of $Y_{2D} = 350\pm20$~N/m. The error on the Young's modulus is given by the fitting error.
\label{S_fig_06}}
\end{figure*}

\begin{figure*}
\begin{center}
\includegraphics[width=102.5mm]{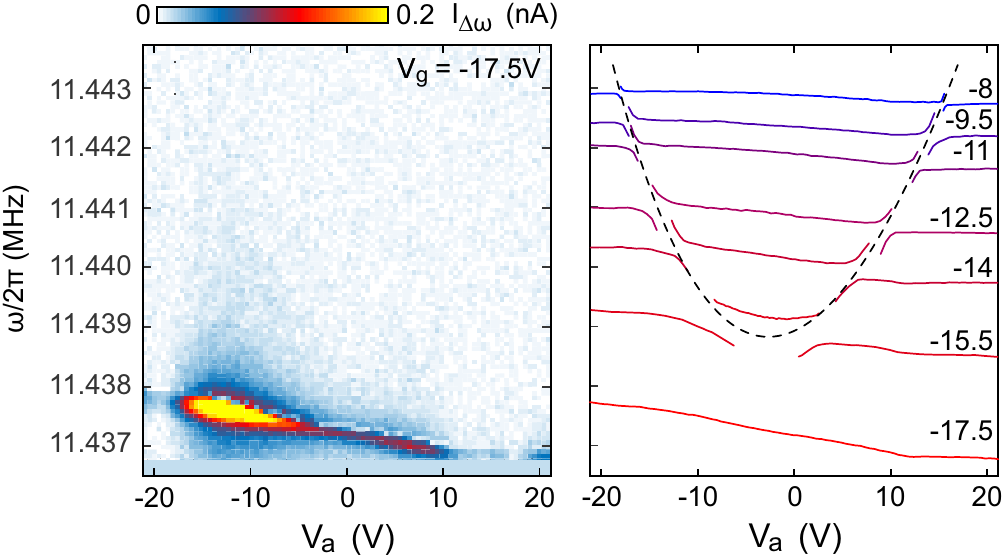}
\end{center}
\caption{
The left panel shows the raw data of the down-mixing current $I_{\Delta\omega}$ as a function of $\omega/2\pi$ and $V_\text{a}$ for $V_\text{g} = -17.5$~V (black label). This data set does not reveal any avoided crossings. The colored lines in the right panel indicate the extracted resonance frequencies for different fixed $V_\text{g}$ (black labels), including the extracted frequencies at $V_\text{g} = -17.5$~V. The dashed black line traces the dependence of the avoided crossings on $V_\text{a}$.
\label{S_fig_08}}
\end{figure*}

\renewcommand{\figurename}{Supplementary Table}
\setcounter{figure}{0}

\begin{figure*}
\begin{center}
\includegraphics[width=105mm]{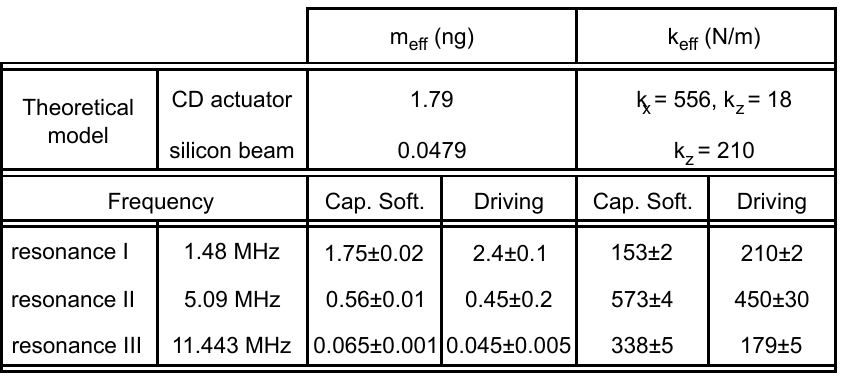}
\end{center}
\caption{Overview of the theoretical and experimentally determined effective masses and spring constants. The theoretical values were obtained from finite element analysis of the studied device. $k_\text{x}$ denotes the spring constant in the $x$-direction and $k_\text{z}$ in the $z$-direction. Experimentally, we determined the effective masses and spring constants (i) from equation~\ref{eq2} in the main text taking into account the capacitive softening (Cap. Soft.) effect and (ii) from the effective drive amplitude $A$ as a function of drive force (Driving). Based on the theoretical masses, we attribute resonances I and II to the comb-drive actuator and resonance III to the suspended silicon beam (see main text).
\label{S_tab_01}}
\end{figure*}

\begin{sidewaysfigure}
\vspace{8cm}
\begin{center}
\includegraphics[width=218mm]{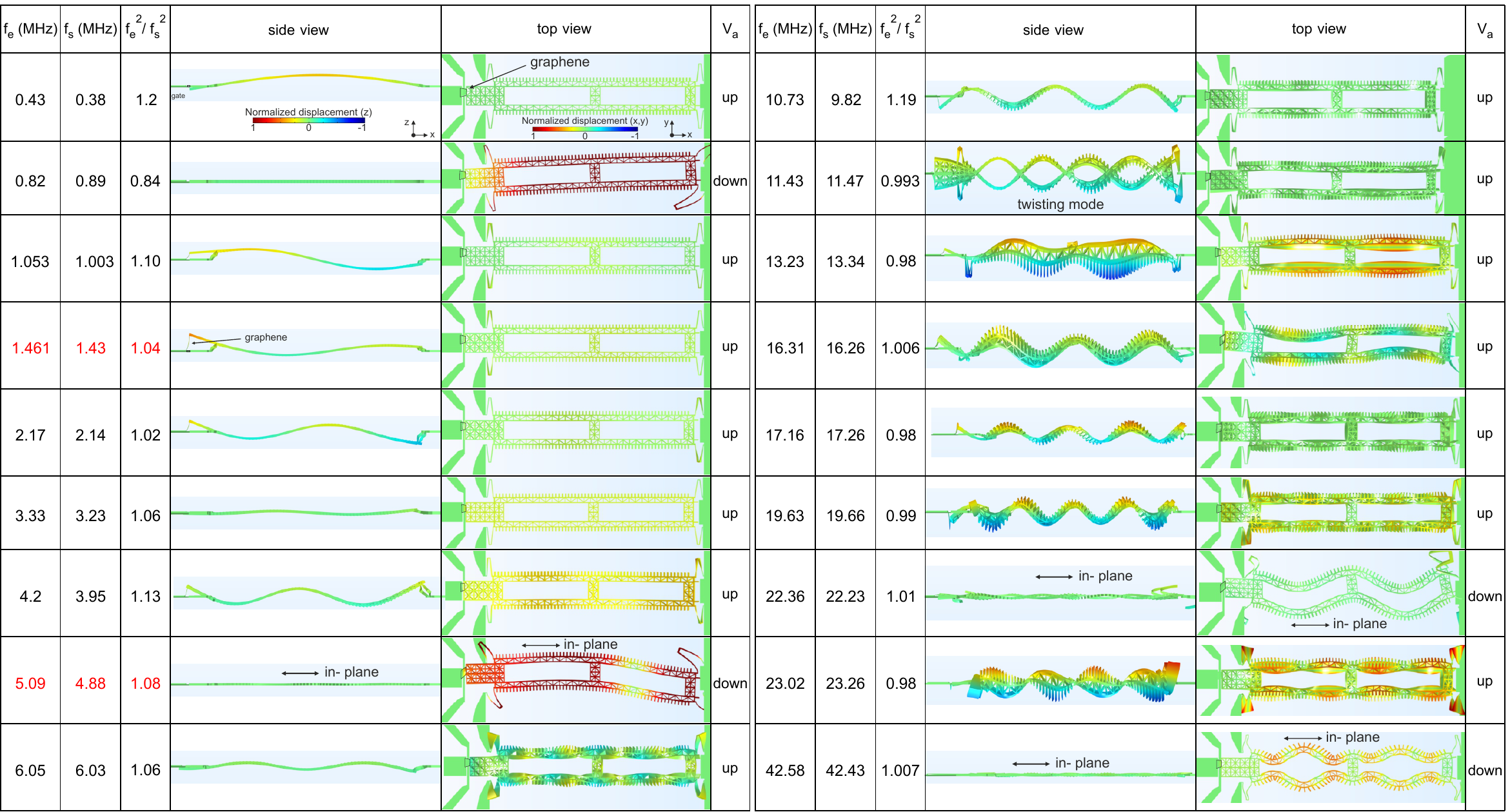}
\end{center}
\caption{
Mode shape overview. The first column gives the experimentally measured frequency $f_\text{e}$ and the second column gives the corresponding frequency $f_\text{s}$ extracted from the finite element simulation. The third column gives the squared ratio between $f_\text{e}$ and $f_\text{s}$. The fourth and fifth column give a side and top view of the mode shape, respectively. Note that the color scale for each figure in the same column is identical such that the amplitudes can be directly compared to one another. The last column specifies the observed tuning with applied potential $V_\text{a}$ to the actuator. All modes tune downwards with applied potential $V_\text{g}$ to the silicon beam underneath the graphene. Resonances I and II defined in the main manuscript at 1.461~MHz and 5.09~MHz, respectively, are highlighted in red. The table contains 16+1 resonances below 25~MHz, just as the experimental down-mixing current in Figure~1f of the main manuscript.
\label{S_tab_02}}
\end{sidewaysfigure}

\end{document}